\documentclass[runningheads,a4paper]{llncs}

\usepackage{amssymb}
\usepackage{amsmath}
\usepackage{graphicx}
\usepackage{color}
\usepackage{soul}
\usepackage{breqn}
\usepackage[utf8x]{inputenc}
\usepackage{fixltx2e}
\usepackage[nounderscore]{syntax}
\usepackage[boxruled]{algorithm2e}

\usepackage{url}
\urldef{\mailsa}\path|jan@kozusznik.cz|
\newcommand{\keywords}[1]{\par\addvspace\baselineskip
\noindent\keywordname\enspace\ignorespaces#1}

\begin{document}

\mainmatter  % start of an individual contribution
\bibliographystyle{splncs03}
\newcommand{\U}{\mathcal{U}}
\newcommand{\G}{\mathcal{G}}

\newcommand{\Links}{\mathcal{L}}
\newcommand{\States}{\mathcal{S}tates}
\newcommand{\links}{L}
\newcommand{\mstates}{state}
\newcommand{\mrelevance}{\rho}

\newcommand{\relevantVertices}{rel_{V}}
\newcommand{\relevantEdges}{rel_{E}}
\newcommand{\prohibited}{prohibited}
\newcommand{\defr}[1]{\mathcal{D} (#1)}
\newcommand{\emp}[1]{\textit{#1}}
\newcommand{\fig}[3]{
	\begin{figure}[ht] 
	\centering 
	\includegraphics[width=0.49\textwidth]{#1}
\caption{#2}
\label{#3}
\end{figure}}

\newcommand{\lame}[2]{\left(\lambda\hspace{1mm}#1\right)\left(#2\right)}
\newcommand{\deltaDel}[1]{{#1_{\Delta}}_{del}}
\newcommand{\deltaCrt}[1]{{#1_{\Delta}}_{crt}}
\newcommand{\deltaUpd}[1]{{#1_{\Delta}}_{udp}}
\newcommand{\alignedm}[1]{\begin{aligned}#1\end{aligned}}

%not used equation
\ifx true false
\begin{equation}
\begin{aligned}
IsEdgeInPath = \lame{e,p}{p=i_1,i_2\ldots i_n \wedge \right.\\
\left.{}\wedge i_j=e\wedge j\bmod{2} = 1}
\end{aligned}
\label{eq:IsEdgeInPath}
\end{equation}

\begin{algorithm}[ht]
\DontPrintSemicolon
\LinesNumberedHidden
${O_{\Delta}}_{crt} \gets {O_{\Delta}}_{crt} \cup \{i \in Identity\mid \exists c\colon c \in {O_{\Delta}}_{crt}\cap Contact \wedge c.contactIdentity = i\}$\;
\caption{Creation of contact}
\label{alg:creationOfContact}
\end{algorithm}
The second issue is solved in a client part whilst it updates its storage by adding this extra operation(see algorithm \ref{alg:deletionOfContact}).
\begin{algorithm}[!ht]
\DontPrintSemicolon
\LinesNumberedHidden
\SetKw{Del}{delete}
\ForAll{$c \in {O_{\Delta}}_{del} \cap Contact$}{
	\Del c.contactIdentity
}
\caption{Deletion of contact}
\label{alg:deletionOfContact}
\end{algorithm}

The same patterns are used for participation and referred event in case of a creation of a participation(see algorithm \ref{alg:creationOfParticipation}) and also in case of a deletion of a participation(see algorithm \ref{alg:deletionOfParticipation}).

\begin{algorithm}[!ht]
\DontPrintSemicolon
\LinesNumberedHidden
${O_{\Delta}}_{crt} \gets {O_{\Delta}}_{crt} \cup \{e \in Event\mid \exists p\colon p \in {O_{\Delta}}_{crt}\cap Participation \wedge p.Event = e\}$\;
\caption{Creation of participation}
\label{alg:creationOfParticipation}
\end{algorithm}

\begin{algorithm}[!ht]
\DontPrintSemicolon
\LinesNumberedHidden
\SetKw{Del}{delete}
\ForAll{$p \in {O_{\Delta}}_{del} \cap \{p \in Participation\mid p.Identity = i_u\}$}{
	\Del p.Event
}
\caption{Deletion of participation}
\label{alg:deletionOfParticipation}
\end{algorithm}
We can use this solution because a specific contact identity respectively an specific event is included into relevant data only with referencing contact respectively participation. The more general solution is to change the way we record that some link has been created or deleted. It will be recorded separately independently on an owning object and it will be another task in the future research.

There are two hosts $A$ and $B$ and every holds its set $S_A$ respectively $S_B$ of b-bit numbers. The goal of the problem is for $A$ and $B$ to each compute $S_A \cap S_B$ with a minimum communication\cite{minsky_practicalset_2002}. Let's ignore that our problem works with structured data. The most significant difference is that the mentioned solution has sets $S_A$ and $S_B$ as input whilst the solution presented in our paper has as input a set $S_W$. This set represents all data contained in a server part of the system. The set $S_A \subset S_W$ has to be computed during solution of the problem and it represents relevant data mentioned earlier.
\fi
\title{Mobile Device Synchronization with Central Database based on Data Relevance}

\titlerunning{Mobile Device Sync. with Central
Database based on Data Relevance}

\author{Jan~Ko\v{z}usznik}

\institute{V\v{S}B - Technical University Ostrava\\
17. listopadu, Ostrava, Czech Republic\\
\mailsa\\
\url{http://www.kozusznik.cz}}


\toctitle{Lecture Notes in Computer Science}
\tocauthor{Authors' Instructions}
\maketitle

\begin{abstract}
Distributed applications are broadly used due the existence of mobile devices as are mobile phones, tablets and chrome books. They are often based on an architecture client-server. A server part contains a central storage where all application data are stored. A specific user accesses data using a client part of an application. The client part of the application can store data in its local storage to enable an offline mode. In the past, we were developing such a client-server software system and we realized that it doesn't exists a general way how to define, search and provide into a client part only those data that are important or relevant to a particular user. The aim of this article is a definition of the problem and a creation of its general solution, not how to achieve the most effective and complete data synchronization. Complete synchronization of data that has been already theoretically and practically solved on a general level and it is impractical for our case.
\end{abstract}

\numberwithin{equation}{section}

\section{State Of The Art}\label{state-of-the-art}
A synchronization of mobile devices is mentioned from a different point of view in existing and already publicised researches.\\
Overview of commonly used synchronization protocols for PDAs is in the paper\cite{agarwal_scalability_2002}. Mentioned protocols were being used for synchronization of full PDA storage between many devices used only by one specific user. Compared protocols are Hot Sync used by Palm OS; Intellisync used by company Intellisync acquired by Nokia in 2006; SyncML that is open industry initiative supported by many companies. Last one protocol is CPIsync that is an abbreviation to Characteristic Polynomial Interpolation Synchronization. The protocol is based on algebraic principles and it is described in \cite{minsky_set_2003} or \cite{trachtenberg_fast_2002}. In the paper, a problem of a synchronization is called as the set reconciliation problem. Presented principle for own synchronisation promises better results then a principle of timestamps that is used in our paper and we will try to use it with our solution in the future.\\
It is described a new architecture in an another paper\cite{brodt_mobile_2011} that promises interoperability between different application at the data management level. Beside data store, architecture contains also meta-data store where is used RDF\cite{needleman_rdf:_2001}.
An another interesting approach is offered in \cite{kaushik_smartparcel:_2013}. There is described an architecture that provides a collaborative data sharing among spatial-temporally co-existing mobile devices. An aim of the approach is to minimizing an overall network usage and a traffic.

\section{Introduction}
In this paper, we present the problem that was solved during development of the system that will be used for a communication between users who are connected in some internet social network e.g. Twitter, LinkedIn or Google+. Users work with the system on their mobile devices with temporary Internet access.\\
Description of the problem is in section~\ref{sec:problemStatement}. Section~\ref{sec:definitions} contains formal definitions of basic concepts used in this paper. There is a more detailed description of the system in the section \ref{sec:systemDescription}. Generalization of a selection of relevant data is in the section~\ref{sec:selectionRelevant}. How is a synchronization performed according to selected relevant data shows section~\ref{sec:principleSync}. In the section \ref{sec:Discusion}, known existing issues of the solution are discussed. The section \ref{sec:Conclusion} contains a conclusion.
\section{Problem definition}\label{sec:problemStatement}
The architectural style of the system is client-server\cite{PfleegerSE2009}. The server part of the system contains all the information - such as user profiles, their contacts (connection to another user in some social networks), and more. The server part of the system should be defined marketing term \textit{cloud} (see \cite{rhoton_cloud_2009}). Part of the system that runs on a mobile device is referred to as the client part of the system. Generally, a term data is used for set of objects (entities), their state and connections or links among them (see definition \ref{df:defData}). The client part of the system should contain current information to which the user has permission and are also important for him. Such information is referred to as the relevant information (see definition \ref{df:relevantData}).\\
The term data in our paper is used for the set of all objects (entities), their states and all links between these objects.\\
Timeliness of data means that the data in the client part are identical to relevant data in the server for the user.\\
The main objective of this paper is not a description of the problem of synchronization but how to specify the relevant data and how to find modifications for a specific user in these data. These have to be transferred into the client part of the application. Synchronization is based on the principle of timestamps.\\
\section{Basic definition}\label{sec:definitions}
Running object-oriented system is composed of objects that are interconnected by links. 
The links are of some type. Type of a link is called association\cite{blaha_object-oriented_2005}. Objects, their state and links define data of the system.
\begin{definition}\label{df:defData}
Let $O$ is the set of all objects that the system is working with\footnote{Only persistent objects are considered.}. Let $A$ is the set of all associations in system. Let $L \subset O \times O \times A$ is the set of links between objects. We say that there is a link of a specific association ``$a$'' between objects $o_1,o_2\in O$ if and only if $(o_1,o_2,a)\in L$. Let $S$ is the set of all possible states of all objects in the system and let the map ``$\mstates: O \longrightarrow  S$'' exists. The map ``$\mstates$'' returns for a certain object state of the object. There is used term \textbf{system data} for the triple $data=(O, L, \mstates)$ in this paper. If we want to talk about the system data in time $t$ then it is denoted as $(O_t, L_t \ mstates_t)$.
\end{definition}
In object-oriented programming environments, objects are usually defined by the classes\cite{blaha_object-oriented_2005}. Every object is an instance of exactly one class. 
The classes are so mutually disjunctive. Type of a link is expressed as association\cite{blaha_object-oriented_2005}. Each association is defined by two classes and two roles in which objects of the classes are found. These classes define permitted types of objects that can be linked with that type. Roles define names that objects in the link can be referred with.
\begin{definition}\label{df:structureOfData}
Let  $D=(O, L, \mstates)$ are system data. Let $C$ is the set of all classes of all objects in the system. Let $A$ is the set of all associations defined in the system. This set is defined as $A\subset C \times C \times Roles \times Roles$. If it is satisfied the equation \ref{eq:structureOfData} then the pair $(C,A)$ is called \textbf{a schema of system data} $D$.\\
This relationship is expressed by the predicate $IsSchemaOf = (\lambda s,d)($schema $s = (C,A)$ is a schema of the system data $d=(O, L, \mstates))$
\end{definition}

\begin{equation}
\begin{alignedat}{1}
C\subset 2^O \wedge &(\forall o:o\in O)((\exists c: c \in C)(o \in c))\wedge(\forall c_1,c_2)(c_1,c_2 \in C \wedge c_1\neq c_2 \\
&\Rightarrow c_1\cap c_2 = \emptyset)\\
{}\wedge (\forall l: &l \in L)((\exists a: a \in A)(l = (o_1, o_2, a)\wedge a = (c_1, c_2, a_p) \\
&\wedge o_1 \in c_1\wedge o_2 \in c_2))
\label{eq:structureOfData}
\end{alignedat}
\end{equation}
If the storage of the server part contains the object $o_i$ and the storage of the client part contains the object $o_{ii}$ and both objects represent the same entity then these objects refer to a single item $o$ from the set of objects $O$ although their states vary in both repositories. We can then write $o_i = o = o_{ii}\wedge o \in O$.
\begin{definition}\label{df:subsetOfData}
Let $d_1$ and $d_2$ are various system data where $d_1=~(O_1,\links_1,\mstates_1)$ and $d_2=~(O_2,\links_2,\mstates_2)$. Let $s$ is a schema that meets $IsSchemaOf(s,d_1) \wedge IsSchema(s,d_2)$ and $s = (C,A)$.
We say that $d_2$ are \textbf{sub-data} of $d_1$ if and only if $O_2\subset O_1\wedge\links_2\subset\links_1\cap O_2\times O_2\times L \wedge \mstates_2=\mstates_1\cap O_2\times S$. This is denoted as $d_2\subset d_1$.
\end{definition}
\begin{definition}\label{df:relevantData}
Let $U$ is the set of all system users, $D$ is the set of all possible system data - $D=2^O\times 2^{O\times O\times A}\times 2^{O\times S}$ and let there exist a map $\mrelevance: U\times D \Longrightarrow D$ that meets $(\forall u\colon U) (\forall data \colon D)  \left(\mrelevance(u,data) \subset data\right)$. In this case, the map $\mrelevance$ is called as the \textbf{selection of relevant data} and result of this map for the user $u$ and system data $data_s^u$ is called \textbf{relevant data} in $data_s$ for the user $u$. Let there any relevant data such that it satisfies $data_r=(O_r, \links_r, \mstates_r)$, then we call $O_r$ relevant objects, $\links_r$ relevant links and $\mstates_r$ relevant states. The above mentioned relationship between the user, the system data and the relevant data is expressed by the predicate  $RelevantDataForUser = (\lambda u,d,d_r)(d_r$ are relevant data of system data $d$ for the user $u)$
\end{definition}

\section{Description of the system}\label{sec:systemDescription}
The system allows the user $u_o$ to invite selected friends (other users included between the contacts of the user) to the event organized by the user $u_o$. Invitation to the event for each user is transferred to the mobile device. The invited user can accept this invitation or refuse. Invite status are visible to other participants in the same event. The selected part of the ER model for this system is shown in Figure \ref{fig:data_model}.
\fig{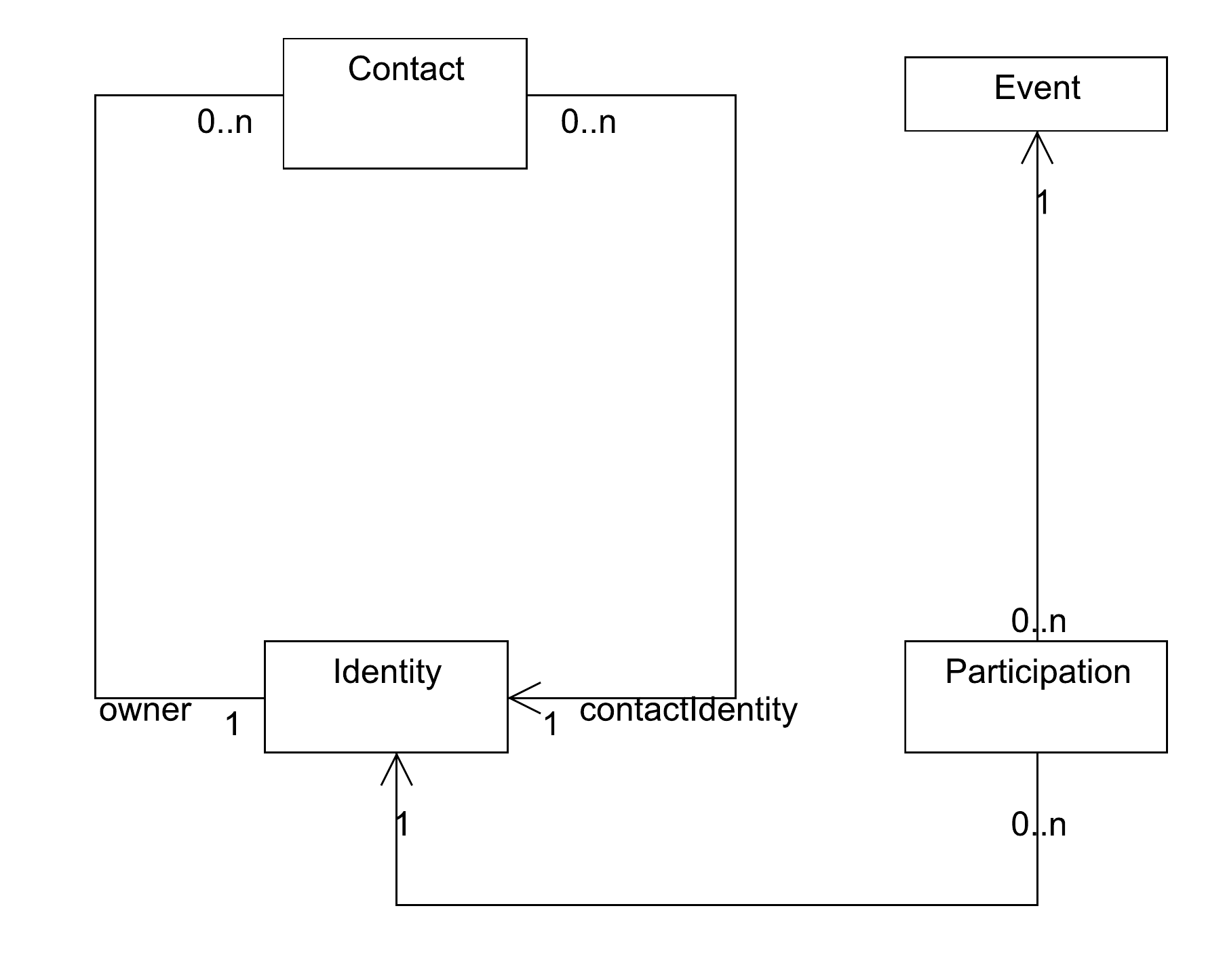}{The object model of the system}{fig:data_model}
Let $u$ is a user that is represented by an object $i_u$ which is an instance of the class Identity. During synchronization is necessary to ensure the timeliness of the information about the user, his contacts and events to which he is invited. Relevant data $data_r^u=(O_r^u, \links_r^u, \mstates_r^u)$ are chosen. Definition \ref{df:relevantData} is general and depends on the structure and meaning that data are actually relevant (as is done selection of relevant data)  and it is specific to each system. The following is a description of the selection of relevant data for our system. A more general description is then in the section \ref{sec:selectionRelevant}.\\
Relevant objects $O_r^u$ for user $u$ can be divided into three basic sets - $O_r^u={O_r^u}_1\cup{O_r^u}_2\cup{O_r^u}_3$. These sets are mutually disjoint and satisfy ${O_r^u}_1\cap{O_r^u}_2\cap{O_r^u}_3=\emptyset$.\\
The first set contains only one item $i_u$ and it is true ${O_r^u}_1=\left\{i_u\right\}$.\\
The second set ${O_r^u}_2$ consists of two subsets ${{O_r^u}_2}_c$ and ${{O_r^u}_2}_i$ and we write ${O_r^u}_2 = {{O_r^u}_2}_c \cup {{O_r^u}_2}_i$. ${{O_r^u}_2}_c$ is the set of contacts of the user $u$ where $i_u$ is in the role of the owner of such contacts (see equation \ ref {eq: contactsOfIdentity}). All identities that are linked to some of these contacts in the role contactIdentity are included in the set ${{O_r^u}_2}_i$ but without any other objects (see equation \ref{eq:identityInContact}).
\begin{align}
{{O_r^u}_2}_c&=\left\{c\in Contact\mid c.owner=i_u\right\}
\label{eq:contactsOfIdentity}\\
{{O_r^u}_2}_i&=\left\{i\in Identity\mid\left(\exists c\colon{{O_r^u}_2}_c\right)
										    \left(c.contactIdentity = i\right)\right\}
\label{eq:identityInContact}
\end{align}
The last set contains instances of the class Participation, which are linked with that identity $i_u$ (\ref{eq:partOfIdent}). Furthermore, this set contains all the events associated with any such object Participation(see equation \ref{eq:eventOfPart}). Finally, this set contains all objects of the class Participation linked to any such event and the identity that $i_u$ has among his contacts. User can only see a participation of his friends. The set does not contain any other elements: ${O_r^u}_3={{O_r^u}_3}_e\cup{{O_r^u}_3}_p\cup{{O_r^u}_3}_p$.\\
\begin{align}
{{O_r^u}_3}_p&=\left\{p\in Participation\mid p.Identity = i_u\right\}
\label{eq:partOfIdent}\\
{{O_r^u}_3}_e&=\left\{e\in Event\mid\left(\exists p\colon{{O_r^u}_3}_p\right)\left(p.Event=e\right)\right\}
\label{eq:eventOfPart}\\
{{O_r^u}_3}_p&=\left\{ p\in Participation\mid\right.\left(\exists e\colon e\in{{O_r^u}_3}_e\right)\left(p.Event=e\wedge i\in {{O_r^u}_2}_i\right.\nonumber\\
&\qquad\left.\left.{}\wedge p.Participation=i\right)\right\}
\label{eq:partInEvent}
\end{align}
Links are not subject to any restrictions \footnote{This is expected in our system but it is not generally true}. Each link that exists between objects in original data also exists in the relevant data (if both objects are also in the relevant data) - $\links_r^u = \links \cap O_r^u\times O_r^u\times L$. In a similar manner, the relevant state is derived  - $\mstates_r^u=\mstates\cap (O_r^u\times O_r^u)\times S$.

\section{Selection of relevant data}\label{sec:selectionRelevant}
From the just presented problem, we defined a more general way of how to select relevant data. This approach consists of using constraints to objects that reduce the original set to the set of relevant objects. If we look carefully, these constraints can be divided into three groups:
\begin{enumerate}
\item
These constraints define the class to which the object must belong to or a specific instance (on the basis of its identity). This group is denoted as $C_{g1}$.
%\iffalse
\item
These constraints define the states in which objects must be. Their state is given by the value of their attributes. Attributes that implement link to other objects are not included. This group is denoted as $C_{g2}$.
%\fi
\item 
These constraints define how objects are accessible from (see definition \ref{df:objectAccessibility}) objects that are already included in the set of relevant objects because they satisfy one of the constraints of the previous two groups. This group is denoted as $ C_ {g3} $.
\end{enumerate}
The constraint which will include object $i_u$ to the set of relevant objects is representative of the first group. The constraints that will include all the contacts (and referenced identities) or objects of the class Participation for the object $i_u$ to the set of relevant objects are representatives of the third group. In our application, there are no examples of the second group of constraints, but we expect their existence.\\
\ifx true false
A result of the selection changes during time. It depends on a creation of objects, their modifications and deletions and it also depends on changes of links. The last two groups of constraints can be divided orthogonally whether an object satisfies (or not satisfies) constraints during its whole lifetime.\\
An instance of contact is in the set of relevant objects for a user $u$ during its whole lifetime. Whilst an identity, that is referenced be reference contactIdentity is in set of relevant objects only if it is referenced by some contact (of given user $u$).
\fi
\begin{definition}\label{df:graph}
\textbf{Graph with typed edges} is called an ordered triple $\left(V,E,T_e\right)$ where $V$ is the set of vertices, $E$ is the set of edges and $T_e$ is the set of types of edges. $E$ is defined as the set of pairs $\left(e,t_e\right)$ where $e$ is a pair of vertices and $t_e$ is the type of edge so shortly we write $E \subset \left\{\left(v_i, v_{ii}\right)\mid v_i,v_{ii}\in V\right\}\times T_e$. Our definition extends generally used definition of the graph(see \cite{GouldGraph2012}) of the type that is assigned to an edge - $t_e$.
\end{definition}
\begin{definition}\label{df:graphConstruction}
Let $d=\left(O, \links, \mstates\right)$ are the system data and $s=(C,A)$ is the schema. Graph with typed edges $g=(V, E, T_e)$ is called \textbf{graph constructed for system data} if and only if $O=V \wedge T_e=A\wedge \left(\forall l: l \in \links \right)(l = (o_1,o_2, a) \wedge ((o_1,o_2),a) \in E) \wedge (\forall e:e\in E)(e=((e_1,e_2),t_e) \wedge (e_1,e_2,t_e)\in \links$.  The relationship between system and graph constructed for them is expressed by the predicate $GraphForSystemData = (\lambda g,d)(d$ are system data and $g$ is the graph constructed for them$)$.
\end{definition}
Specification of paths that define whether objects (or edges) are included among the relevant objects is based on a simple grammar:
\begin{grammar}
<Sp> ::= <Direct_set_definition>  \alt <Sp>.<Te>
\end{grammar}
$<$Sp$>$ specifies a set of paths. $<$Direct\_set\_definition$>$  specifies only those paths that have a length of zero (containing only one vertex). 
These paths contain only one vertex, which meets certain conditions. We omit the definition of grammatical rules of that non-terminal. Alternative $<$Sp$>$.$<$Te$>$ derives a new set of paths. \\
Let $d=(O,\links,\mstates)$ are system data with schema $s=(C,A)$ and let $g=(V,E,A)$ is a graph that satisfies $GraphForSystemData(g,d)$.
Let $Path_0$ is an expression written using our defined grammar. Let $S_n$ is the set of all paths defined by this expression then an expression $Path_0$.$T_e$ defines a new set of paths $S_{n+1}$. The definition of this set is in \ref{eq:SetOfPath}.
\begin{equation}\label{eq:SetOfPath}
\begin{split}
S_{n+1}&=\{p'\mid p\in S_n \wedge EndVertex(p,v_n)\wedge p'=p,e,v_{n+1} \\
&\qquad\wedge IsPath(p',G)\wedge IsInRole(g,v_{n+1}, e, T_e) \}\\
&\cup S_n \setminus \{p\mid p \in S_n \wedge p'\in S_{n+1} \wedge IsSubPathOf(p,p',G)\}
\end{split}
\end{equation}
The definition of the used predicates follows. $EndVertex$ indicates that the vertex is the end vertex of the path(see \ref{eq:endVertex}). $IsInRole$ indicates that the vertex has a role in the link, which corresponds with a certain edge (see \ref{eq:IsInRole}). $IsPath$ indicates that a given sequence of vertices and edges is a path (see \ref{eq:IsPath}).The predicate $IsSubPath$ indicates that a path is `` prefix'' other paths in the graph (see \ref{eq:IsSubPath}).
\begin{equation}\label{eq:endVertex}
\begin{split}
EndVertex = \left(\lambda p,v\right)&\left(\mbox{p is path }v_1e_1v_2e_2...e_{n-1}v_n\mbox{for }n \geq 0 \wedge v = v_n\right)
\end{split}
\end{equation}
\begin{equation}\label{eq:IsInRole}
\begin{split}
IsInRole=\left(\lambda G,v,e,r\right)  &\left(G=\left(V,E,T_e\right ) \vphantom{\left\{v,v'\right\}}\right.\wedge \left\{v,v'\right\} \subset V\wedge e \in E  \\
&{}\wedge \left\{c,c'\right\} \subset C\wedge v \in c \wedge v'\in c'\wedge a\in T_e\\
&\wedge \left(a = \left(c,c', \left(r,r'\right)\right)\right.\wedge \left(\left(v,v'\right),a\right) \in E\\
&\quad\vee{}a = \left(c',c, \left(r',r\right)\right)\wedge \left.\left(\left(v',v\right),a\right) \in E\right)
\end{split}
\end{equation}
\begin{equation}
\begin{aligned}
IsPath = (\lambda p,G)&(p = v_1,e_1,v_2,e_2\ldots e_{n-1},v_n 
\wedge G=(V,E,T_e)\\
&{}\wedge (\forall i: i \in [1,n-1] \cup Z)
(e_i = ((v_i,v_{i+1}),t)\wedge t\in T_e\\
&\quad{}\wedge \left\{v_i, v_{i+1}\right\} \subset V \wedge e_i \in E)\\
&{}\wedge (\forall i,j: i,j \in [1,n] \cup Z)(i\neq j \Rightarrow v_i\neq v_j)\\
&{}\wedge (\forall i,j: i,j \in [1,n-1] \cup Z)(i\neq j \Rightarrow e_i\neq e_j)
\end{aligned}
\label{eq:IsPath}
\end{equation}
\begin{equation}
\begin{split}
IsSubPath=(\lambda p,p',G)&(IsPath(p,G) \wedge IsPath(p',G)\\
                &{}\wedge p'=p,e_i,v_{i+1},e_{i+1} \ldots e_{n-1},v_n)
\end{split}
\label{eq:IsSubPath}
\end{equation}

\begin{definition}\label{df:objectAccessibility}
Let $g$ is a graph with typed edges, $p$ is a path in the graph, $v$ is a vertex in the graph and $e$ is an edge in the graph. We say that the vertex $v$ is on the path $p$ if and only if $p=v_0e_0v_1e_i...e_{n-1}v_n \wedge v = v_i \wedge i \in \{z\mid z\in\mathbb{Z} \wedge 0 \leq z \wedge z \leq n\}$. This relationship is expressed by the predicate $IsInPath(v,p)$. We say that the edge $e$ is on the path $p$ if and only if $p=v_0e_0v_1e_i...e_{n-1}v_n \wedge e = e_i \wedge i \in \{z\mid z\in\mathbb{Z} \wedge 0 \leq z \wedge z < n\}$. This relationship is expressed by the predicate $IsInPath(e,g)$.
\end{definition}
Let $d$ are the system data with the schema $s$. Relevant data denoted $d_r = (O_r,\links_r,\mstates_r)$. Let $Es_p $ is the set of expressions defined by the above mentioned grammar and $S_p$ is the set of all paths in the data generated $d$ using the expressions of $Es_p$. Sets are constructed based on the equations \ref{eq:subsetOOfRelevant}, \ref{eq:subsetLOfRelevant} and \ref{eq:subsetSOfRelevant}.
\begin{equation}\label{eq:subsetOOfRelevant}
\begin{split}
O_r =\{o\mid o \in O \wedge IsInPath(o,path)
     \wedge path \in S_p\}
\end{split}
\end{equation}
\begin{equation}\label{eq:subsetLOfRelevant}
\begin{split}
\links_r=\{l\mid l \in \links \wedge IsInPath(l,path)
         \wedge path \in S_p\}
\end{split}
\end{equation}
\begin{equation}\label{eq:subsetSOfRelevant}
\begin{split}
\mstates_r = \mstates \cap (O_r\times\links_r\times\States)\\
\end{split}
\end{equation}
Expressions that define the paths for the selection of relevant data for the user $u$ in our system are as follows:
\begin{itemize}
\item$\{id_u\}$.Contact.contactIdentity
\item$\{id_u\}$.Participation.Event.Participation.Identity
\end{itemize}
where $id_u$ is the identity of the user $u$.\\
We note that the specified expressions generate a slightly different set of relevant data than was defined in \ref{sec:systemDescription}. The second term includes the identity to a set of relevant data regardless of whether it is the user contacts or not. This problem will be solved in the future with a modest extension of the grammar, which allows you to define that an object in the path must already be included in any other path.
\section{The principle of synchronization relevant data}\label{sec:principleSync}
At the beginning, we will describe the synchronization, which ensures that the client part has the current data. This part runs on a device that belongs to a certain user. Any changes in the repository of this section is immediately transferred to the server part of the system. This ensures timeliness of its data. The client part is periodically requesting the server part for the changes since the last synchronization of the time $t_ {ls}$. The general algorithm of the searching of changes in the relevant stored in the server part is described in \ref{alg:generalCloudSync}. The inputs of the algorithm are:
\begin{itemize}
	\item $t_{ls}$ - last time of the synchronization
	\item $u$ - the user
	\item $data=(O, \links, \mstates)$ - all data stored in the server part of the system,
	\item $data_{t_{ls}}=(O_{t_{ls}}, \links_{t_{ls}}, \mstates_{t_{ls}})$ - all data stored in the server part of the system in time ${t_{ls}}$.
\end{itemize}
and produces:
\begin{itemize}
	\item ${O_{\Delta}}_{crt} \subset O$ - set of objects to be created in the client part,
	\item ${O_{\Delta}}_{upd} \subset O$ - set of objects to be updated in the client part,
	\item ${O_{\Delta}}_{del} \subset O_{t_{ls}}$ - set of objects to be deleted in the client part,
	\item ${\links_{\Delta}}_{crt} \subset \links$ - set of links to be created in the client part,
	\item ${\links_{\Delta}}_{del} \subset \links$ - set of link to be deleted in the client part,
	
	\item $\mstates`\subset O\times \States$ - current status of newly created or modified objects.
	\item $t_{cs}$ - the time synchronization.
\end{itemize}
\begin{algorithm}[h]
\SetAlgoLined
\DontPrintSemicolon
	$(O_r, \links_r, \mstates_r) \gets \mrelevance(data,u)$\;
	$({O_r}_{t_{ls}}, {\links_r}_{t_{ls}}, {\mstates_r}_{t_{ls}}) \gets \mrelevance(data_{t_{ls}},u)$\;
	$\deltaDel{O}\gets\deltaDel{GetSet}({O_r}_{t_{ls}}, O_r)$\;
	$\deltaCrt{O}\gets\deltaCrt{GetSet}({O_r}_{t_{ls}}, O_r)$\;
	
	$\deltaUpd{O}\gets\deltaUpd{GetSet}({O_r}_{t_{ls}},O_r,{\mstates_r}_{t_{ls}},\mstates_r)$\;
	
	$\deltaDel{\links}\gets\deltaDel{GetSet}({\links_r}_{t_{ls}}, \links_r)$\;
	$\deltaCrt{\links}\gets\deltaCrt{GetSet}({\links_r}_{t_{ls}}, \links_r)$\;
	
	$\mstates`\gets\mstates_r \cap ({O_{\Delta}}_{create} \cup {O_{\Delta}}_{update}) \times \States$\;
	$t_{cs}\gets$ current time\;
\caption{General algorithm of the searching of changes in the server part of the system}
\label{alg:generalCloudSync}
\end{algorithm}
Used construction are defined in \ref{eq:GetSetCrt}, \ref{eq:GetSetDel} and \ref{eq:GetSetUpd}.
\begin{equation}\label{eq:GetSetCrt}
\deltaCrt{GetSet}= \left(\lambda S_l,S\right)\left\{i \mid i \notin S_l \wedge i \in S\right\}
\end{equation}
\begin{equation}\label{eq:GetSetDel}
\deltaDel{GetSet}= \left(\lambda S_l,S\right)\left\{i \mid i \in S_l \wedge i \notin S\right\}
\end{equation}
\begin{equation}\label{eq:GetSetUpd}
\begin{split}
\deltaUpd{GetSet}&=\left(\lambda S_l,S,state_l, state\right)\left\{i\mid i \in S\setminus \deltaCrt{GetSet}\left(S_l,S\right) \right.\\
&\left.\qquad{}\wedge state_l\left(i\right) \neq state\left(i\right) \right\}
\end{split}
\end{equation}
The client part of the application updates its local storage base on the output of the previous algorithm. The described algorithm is simple to describe, but difficult to implement. It requires trace the history of each change because it refers to the system data at the time of the last synchronization $t_{ls}$. \\
Instead of keeping a complete history, we use \emp{timestamps}. The timestamp is an integer that is generated for a group of activities performed at one point or better during one transaction~\cite{BernsteinPrinciples2009} and it is assigned to the actions. Let \textit{act} be a same action. We refer to its timestamp as $ts_{act}$. Let $a_1, a_2$ are some actions, ${ts}_{a_1}, {ts}_{a_2}$ are the timestamps and $Tr_1,Tr_2$ are transactions in which these actions took place. The values ​​of these timestamps satisfy the following conditions:
\begin{itemize}
	\item ${ts}_{a_1}={ts}_{a_2}$ is valid if and only if $Tr_1=Tr_2$. 
	\item ${ts}_{a_1}<{ts}_{a_2}$ is valid if and only if $Tr_1\neq Tr_2$ and the transaction $Tr_1$ has been finished before $Tr_2$\footnote{We assume that two different transactions cannot finish at the same time}.
\end{itemize}
In the server part of the system, it is recorded for each performed action (create, update, delete)  the type of the event, object identifier and the timestamp of the event. For each object, it can exists at most three different time stamps. In the case of repeated modifications of the same object, it is always recorded timestamp only the last action.The system also records the creation and deletion of links between objects, such as information about related objects and type of link (association). Let $At=\{create, update, delete\}$ is the set of action types then we define the map $ts: O\cup\links\times At  \longrightarrow Integer\cup\{None\}$ which returns for a given type and object recorded timestamp or $None$.\\
A modified algorithm for changes finding in the server repository based on the time stamp of the last synchronization - $ts_{ls}$ - is described in \ref{alg:timestampCloudSync}. This algorithm produces timestamp of the performed synchronization $ts_{cs}$. This timestamp is the maximum value of all timestamps of objects that are transferred during the synchronization.\\
In the algorithm, we use the following construction:
\begin{itemize}
	\item predicates $IsLink$ or $IsObject$ returns true if the element is a link or an object. This can be resolved from logs, in which the actions $create, update, delete$ with a time stamp are recorded.
	\item predicate $HasModification$ - that is $true$ if and only if the given path $p$ contains a created, updated or deleted object or link with a newer timestamp than the $ts_{ls}$. This predicate is defined by equation \ref{eq:HasModification}.
	\item predicate $IsInPath$ -which is $true$ if and only if the object or link is in the path $p$ (see an equation \ref{eq:IsInPath}).
	\item $IndexOfFirstCreatedElement$ - smallest index of any element in the path $p$ which is of type $T$ and was created after the time $t$ (see an equation \ref{eq:FirstModified}). In this equation, there is used an predicate $IsIndexOf$ (see an equation \ref{eq:IsIndexOf}).
\end{itemize}

\begin{algorithm}[!ht]
\DontPrintSemicolon
\LinesNumberedHidden
$ts_{cs} \gets ts_{ls}$\;
$\deltaCrt{\links}\gets \emptyset$\;
$\deltaCrt{O}\gets \emptyset$\;
$\deltaUpd{O}\gets \emptyset$\;
$\deltaDel{O}\gets \emptyset$\;

\ForAll{$p \in Paths \wedge HasModification\left(p,ts_{ls}\right)$}{
	$\alignedm{
	\deltaCrt{\links} \gets \{l\in\links\mid\mbox{}IsInPath(l,p)\wedge ts(l,create) > ts_{ls}\}\cup\deltaCrt{\links}}$\;
	$\alignedm{
	\deltaCrt{O}\gets\{o\in O\mid\mbox{}&IsInPath(o,p)\wedge ts(o,create) > ts_{ls}\}\cup\deltaCrt{O}}$\;
	$\alignedm{
	\deltaUpd{O} \gets \{o\in O\mid IsInPath(o,p)
	                   \wedge ts(o,update) > ts_{ls}\}
											\cup\deltaUpd{O}}$\;
	$i_l\gets IndexOfFirstCreatedElement(p,ts_{ls},IsEdge)$\;
$\alignedm{
	\deltaCrt{O} \gets\{o\in O\mid IsInPath(o,p)
                    \wedge IndexOf(o,p,i,G) \wedge i\geq i_l\}
                    \cup\deltaCrt{O}}$\;
$\alignedm{
	\deltaCrt{L} \gets\{l\in L\mid IsInPath(l,p)
                    \wedge IndexOf(l,p,i,G) \wedge i\geq i_l\}
                    \cup\deltaCrt{L}}$\;										
	
	$\deltaUpd{O}\gets\deltaUpd{O}\setminus\deltaCrt{O}$\;
}
$\alignedm{
	\deltaDel{O}\gets\{o:IsObject\mid ts(o,delete) \neq None
                   \wedge ts(o,delete) > ts_{ls}\}
									 \cup\deltaDel{O}}$\;
$\alignedm{
	\deltaDel{\links}\gets \{l:IsLink\mid\mbox{}ts(l,delete) \neq None
	                       \wedge ts(l,delete) > ts_{ls}\}
                         \cup\deltaDel{O}}$\;
$\alignedm{
	ts_{cs} \gets max(&\{ts(o,delete)\mid o \in \deltaDel{O}\}
	                  \cup\{ts(o,update)\mid o\in\deltaUpd{O}\}\\
										&\cup\{ts(o,create)\mid o\in\deltaCrt{O}\}
										\cup\{ts_{cs}\})}$\;
\caption{A mobile synchronization}
\label{alg:timestampCloudSync}
\end{algorithm}
\begin{equation}
\begin{aligned}
HasModification = (\lambda p,t)&(p = e_1,e_2,e_3\ldots e_{n-1},e_n \\
                               &{}\wedge a \in \{create,update,delete\}\\
															 &{}\wedge i\in [1,n]\cap Z\wedge ts(e_i,a) \neq None\\
															 &{}\wedge ts(e_i,a) > t
\end{aligned}
\label{eq:HasModification}
\end{equation}
\begin{equation}
\begin{aligned}
IsInPath = (\lambda i,p)&(p = i_1,i_2,i_3\ldots i_{n-1},i_n \wedge j \in [1,n]\wedge i_j = i
\end{aligned}
\label{eq:IsInPath}
\end{equation}
\begin{equation}
\begin{aligned}
&IndexOfFirstCreatedElement = \\
&\quad\lame{p,ts,T}{min(\{i \in Z \mid T(e)\right.
\wedge IsInPath(e,p)
\wedge IsIndexOf(e,p,i)\\
&\left.\qquad\qquad\qquad\qquad\qquad{}\wedge ts(e,create) \neq None\wedge ts(e,create) > ts \})}
\end{aligned}
\label{eq:FirstModified}
\end{equation}
\begin{equation}
\begin{aligned}
IsIndexOf=(\lambda p,e,j)(p = i_1,i_2,i_3\ldots i_{n-1},i_n\wedge e=i_j)
\end{aligned}
\label{eq:IsIndexOf}
\end{equation}
The client part of the application again updates its local storage base on the output of the previous algorithm. All relevant data are represented by a set $Paths$ that contains all paths. It is defined by set of expression according to the specific grammar. They are selected only those paths that contain the newly created object, updated objects or created link (definition of  predicate $HasModification$ is in the equation \ref{eq:HasModification}). In the beginning, the newly newly created links in a selected path $p$ are included in a set of newly created links $\deltaCrt{\links}$.\\
Likewise, newly created and updated objects in the path $p$ are also included to the sets $\deltaCrt{O}$ and $\deltaUpd{O}$.\\
It is obtained the smallest index of a newly created link in the path. With all of the objects and links that are in the path and they have greater index, it will then treated as a newly created, because they could be unavailable in the previous synchronization.\\
Duplicate objects contained in the set $\deltaCrt{O}$ and the set $\deltaUpd{O}$ are from the set $\deltaUpd{O}$ removed and they are treated as newly created.\\
The set $\deltaDel{O}$ and $\deltaDel{\links}$ contain all the objects and links (their identifiers) that have been deleted since the last synchronization, regardless of whether they are relevant or not. The reason is the server part does not keep information about objects removed, but it keeps only its identification and a time stamp of a removing. The reason is the server part does not keep information about removed objects, but it keeps only its identification and a time stamp of a removing.\\
\section{Discussion of the algorithm}\label{sec:Discusion}
The presented algorithm has several weaknesses known to us. The first of these is that each  removed object or link is transferred to all other mobile devices. This problem is not serious because of the small number of these actions (deletion), because it is not often the case, that would be something deleted.  However, it should be solved more generally in the future.\\
Another shortcoming is again connected with action deletion. Consider the following scenario. The user is in contact with a different identity (representing another user). This contact is deleted but the referenced identity remains still exist. This contact deletion is propagated from the server storage, but referenced identity will not be deleted, even though it is possible that a given user does not have other contact info for this identity. The solution is to check whether the identity of a certain contact is still accesible and delete it if it is not. It is not too general and effective solution and we will work on any better. Briefly, we will register for each object, the number of path, which makes the object in the set of relevant data. During the termination of a path, then this value will be decreased and when it reaches zero, objects can be removed.\\
\section{Conclusion}\label{sec:Conclusion}
In this paper, we have described the problem that we encountered during the development of a single system. In this system, data are shared between multiple users. Subsequently they are  transferred to the mobile devices of users with which they are shared. We want to ensure timeliness of data in repositories of the client part. However, we want to transfer only data that are relevant to the owner of the device.\\
Presented solution has some issues associated with deleting objects and links. In the future, we want to address these issues in a general way.\\
In the future, we also want to test using the synchronization mechanism based on the characteristic polynomial(see \cite{minsky_set_2003}) in our approach.\\

%\bibliography{references}

\end{document}